\documentclass[draft,showpacs,showkeys,eqsecnum,nofootinbib,aps]{revtex4}

\def\beq{\begin{equation}}
\def\eeq{\end{equation}}
\def\bea{\begin{eqnarray}}
\def\eea{\end{eqnarray}}
\def\nn{\nonumber}

\def\pa{\partial}

\def\a{\alpha}
\def\b{\beta}

\begin{document}
\title{De Rham cohomology of $CP^{1}$ model with Hopf term}
\author{Soon-Tae Hong}
\email{soonhong@ewha.ac.kr}
\affiliation{Research Institute for Basic Sciences and Department of Science
Education, Ewha Womans University, Seoul 120-750, Korea}
\date{\today}%
\begin{abstract}
We investigate the $CP^{1}$ model possessing the Hopf term which respects the second class constraints and admits the well defined BRST operator $Q$. Using the operator $Q$, we explicitly construct its de Rham cohomology group by deriving the ensuing quotient group via both the collections of all $Q$-closed and $Q$-exact ghost number $p$-forms. Moreover, we study the $CP^{1}$ model without the Hopf term to evaluate the ensuing effect of the Hopf term on the cohomology group. We find that the Hopf term effects on the de Rham cohomology originate from the non-compact Hilbert space modified by this Hopf term. 
\end{abstract}
\maketitle


The $CP^{1}$ model is a complex extension of the Heisenberg
$O(3)$ model.  The $O(3)$ model emerges in various physical applications from high
energy physics~\cite{niemi} to condensed matter physics~\cite{babaev}. Recently,
there have been some progresses in knotted solitons~\cite{niemi2}. The action of the 
$CP^{1}$ model with the Hopf term possesses a desirable manifest locality since the 
Hopf term has a local integral representation in terms of the physical fields of the 
$CP^{1}$ model~\cite{wilczek82}. Our aim is to investigate the Hopf term effects on de Rham 
cohomology~\cite{derham,derham2,derham3,hong15,hong10}  involved 
in the $CP^{1}$ model possessing the this term. To do this at first we will recapitulate shortly the first class formalism 
of the $CP^{1}$ model without the Hopf term~\cite{hong01}. The $CP^{1}$ model without the Hopf term is defined by the Lagrangian,
\begin{equation}
L_{0}=\int d^{2}x\left[(\pa_{\mu}\xi^{*}_{\alpha})(\pa^{\mu}\xi_{\alpha})
-(\xi^{*}_{\alpha}\pa_{\mu}\xi_{\alpha})(\xi_{\b}\pa^{\mu}\xi^{*}_{\b})\right]. \label{lag}
\end{equation}
Here $\xi_{\a}=(\xi_{1},\xi_{2})$ is a multiplet of complex scalar fields satisfying a constraint
\beq
\omega_{1}=|\xi_{\a}|^{2}-1\approx 0,
\label{omega1}
\eeq
and thus the $CP^{1}$ model becomes a second class constrained Hamiltonian system. 
Exploiting the Lagrangian (\ref{lag}), we find that 
the canonical Hamiltonian is given by
\beq H_{0}=\int
d^{2}x\left[|\pi_{\a}|^{2}+|\pa_{i}\xi_{\alpha}|^{2}
-(\xi^{*}_{\alpha}\pa_{i}\xi_{\alpha})(\xi_{\b}\pa_{i}\xi^{*}_{\b})\right],
\label{canH} \eeq 
where $(\pi_{\a},\pi_{\a}^{*})$ are the canonical momenta conjugate to
the complex scalar fields $(\xi_{\a},\xi_{\a}^{*})$. By implementing the Dirac 
algorithm~\cite{dirac64} we find that, together with Eq. (\ref{omega1}), our Hamiltonian system 
is subject to another second class constraint \beq
\omega_{2}=\xi^{*}_{\a}\pi^{*}_{\a}+\xi_{\a}\pi_{\a}\approx 0. \label{const22} \eeq 
Introducing two canonically conjugate St\"uckelberg fields $(\theta, \pi_{\theta})$ we obtain the 
first class Hamiltonian
\beq
\tilde{H}_{0}=\int
d^{2}x\left[|\pi_{\a}-\frac{1}{2}\xi^{*}_{\a}\pi_{\theta}|^{2}R
+|\partial_{i}\xi_{\a}|\frac{1}{R}
-(\xi^{*}_{\a}\pa_{i}\xi_{\a})(\xi_{\b}\pa_{i}\xi^{*}_{\b})\frac{1}{R^{2}}\right],\nn\\
\label{hct}
\eeq 
where $R=|\xi_{\a}|^{2}/(|\xi_{\a}|^{2}+2\theta)$.


Up to now, we briefly summarize the first class formalism of the $CP^{1}$ model appeared in Ref.~\cite{hong01}. Now 
we are ready to investigate the cohomology group of the $CP^{1}$ model which is our main theme. 
Introducing two canonical sets of
ghost and anti-ghost fields together with auxiliary fields
$(c^{i},\bar{p}_{i})$, $(p^{i}, \bar{c}_{i})$, $(n^{i},b_{i})$ $(i=1,2)$, we define the BRST operator 
for our constraint algebra
\beq
Q = \int {\rm d}^{2}x~(c^{i}\tilde{\omega}_{i}+p^{i}b_{i}),
\label{brstq}
\eeq
from which we find 
\beq
\delta_{Q}\tilde{H}_{0}=0.
\eeq 
Here one notes that $\tilde{\omega}_{i}$ $(i=1,2)$ are the first class constraints of the previous second class ones 
$\omega_{i}$ in Eqs. (\ref{omega1}) and (\ref{const22}).
Moreover, one can readily shows that the BRST charge $Q$ is nilpotent: $Q^{2}=0$.

We choose the unitary gauge with $(\chi^{1},\chi^{2})=(\omega_{1},\omega_{2})$
by selecting the gauge fixing functional 
\beq
\psi = \int {\rm d}^{2}x~(\bar{c}_{i}\chi^{i}+\bar{p}_{i}n^{i}).
\eeq
We then have \beq
\delta_{Q}\delta_{Q}\psi=0, \label{qh} 
\eeq 
which follows from the nilpotentcy of the charge $Q$. Now the gauge fixed BRST invariant Hamiltonian is given by
\bea
H_{eff}&=&\tilde{H}-\delta_{Q}\psi,\nn\\
\tilde{H}&=&\tilde{H}_{0}+\int {\rm d}^{2}x~\left(\frac{1}{2}\pi_{\theta}\tilde{\omega}_{2}-c^{1}\bar{p}_{2}\right).
\eea
Here one notes that in order to guarantee the BRST invariance of $H_{eff}$ we have included in $H_{eff}$ the $Q$-exact 
term, and in $\tilde{H}$ the term possessing $\pi_{\theta}$ and the Faddeev-Popov ghost term~\cite{faddeev67}. 
Moreover, the term $\delta_{Q}\psi$ fixes the particular unitary gauge corresponding to the fixed point 
$(\theta=0,\pi_{\theta}=0)$ in the gauge degrees of freedom associated with two-dimensional manifold described by the 
internal phase space coordinates $(\theta,\pi_{\theta})$, which are two canonically conjugate 
St\"uckelberg fields.

Next, we introduce the BRST operator via
\beq
Q: \alpha_{p}\rightarrow \alpha_{p+1},
\eeq 
where $\alpha_{p}$ is a ghost number $p$-form. We then can construct the $p$-th de Rham cohomology group 
$C^{p}(M,R)$ of the manifold $M$ and the field of real number $R$ with the following quotient 
group~\cite{derham,derham2,derham3,hong15,hong10}
\beq
C^{p}(M,R)=\frac{Z^{p}(M,R)}{B^{p}(M,R)}.
\eeq
Here $Z^{p}(M,R)$ are the collection of all $Q$-closed ghost number $p$-forms $\alpha_{p}$ for which $Q\alpha_{p}=0$, and 
$B^{p}(M,R)$ are the collection of all $Q$-exact ghost number $p$-forms $\alpha_{p}$ for which $\alpha_{p}=Q\alpha_{p-1}$.

The Hamiltonians $\tilde{H}$ and $H_{eff}$ are readily shown to be the ghost number 0-forms, and they are 
$Q$-closed as in the case of $\delta_{Q}\psi$. These Hamiltonians thus can be used to define the $Z^{0}(M,R)$. Here $M$ is the non-compact Hilbert space of the $CP^{1}$ model and $R$ is the real number field. Next, we notice that $\psi$ is the ghost number $(-1)$-form and $Q\psi$ is $Q$-exact ghost number 0-form, so that we can define the $B^{0}(M,R)$.
With these $Z^{0}(M,R)$ and $B^{0}(M,R)$, we construct the 0-th de Rham cohomology group $C^{0}(M,R)$: 
\beq
C^{0}(M,R)=\frac{Z^{0}(M,R)}{B^{0}(M,R)},~~~{\rm for}~CP^{1}~{\rm model}.
\eeq
It is interesting to note that the ghost number 0-form $H_{eff}$ is deformed 
into the other ghost number 0-form $\tilde{H}$. Namely, $H_{eff}$ is homologous to $\tilde{H}$ under the BRST transformation $Q$, or $H_{eff}\sim\tilde{H}$, since $Q\psi=\tilde{H}-H_{eff}$.
These features also appear~\cite{hong15,hong10} even in the 't Hooft-Polyakov monopole~\cite{thooft742,polyakov74}, which is physics-wise different from the $CP^{1}$ model.

Now we generalize the above results to the $CP^{1}$ model with Hopf term. The new Lagrangian is given by
\beq
L_{0}^{H}=L_{0}-\int {\rm d}^{2}x~\frac{\Theta}{8\pi^{2}}\epsilon^{\mu\nu\rho}(\xi_{\a}^{*}\pa_{\mu}\xi_{\a}
-\pa_{\mu}\xi_{\a}^{*}\xi_{\a})\pa_{\nu}\xi_{\b}^{*}\pa_{\rho}\xi_{\b}.
\label{l0h}
\eeq
Here one notes that our physical system with the Hopf term also respects the constraints $\omega_{i}$ $(i=1,2)$ in Eqs. (\ref{omega1}) and (\ref{const22}). Following the same algorithm discussed above, we arrive at the first class Hamiltonian with Hopf 
term~\cite{hong02}
\beq
\tilde{H}^{H}_{0}=\int {\rm d}^{2}x~\left[|\pi_{\a}^{H}-\frac{1}{2}\xi_{\a}^{*}\pi_{\theta}
+\frac{\Theta}{8\pi^{2}R^{2}}\pi_{\a}^{\Theta}|^{2}R+|\pa_{i}\xi_{\alpha}|^{2}\frac{1}{R}
-(\xi^{*}_{\alpha}\pa_{i}\xi_{\alpha})(\xi_{\b}\pa_{i}\xi^{*}_{\b})\frac{1}{R^{2}}\right].
\label{canHh} 
\eeq
Here $\pi_{\a}^{H}$ originates from the canonical momenta conjugate to the complex scalar fields $\xi_{\a}$ associated 
with the $CP^{1}$ Lagrangian with the Hopf term (\ref{l0h}), while $\pi_{\a}^{\Theta}$ is related with the difference between $\pi_{\a}$ and $\pi_{\a}^{H}$. Introducing the ghost and anti-ghost fields together with their auxiliary fields as above, 
we obtain the identities
\bea
H_{eff}^{H}&=&\tilde{H}^{H}-\delta_{Q}\psi,\nn\\
\tilde{H}^{H}&=&\tilde{H}_{0}^{H}+\int {\rm d}^{2}x~
\left(\frac{1}{2}\pi_{\theta}\tilde{\omega}_{2}-c^{1}\bar{p}_{2}\right).
\eea
Resorting to the routine procedure above, we proceed to formulate the cohomology group as follows
\beq
C^{0}(M^{H},R)=\frac{Z^{0}(M^{H},R)}{B^{0}(M^{H},R)},~~~{\rm for}~CP^{1}~{\rm model}~{\rm with}~{\rm Hopf}~{\rm term}.
\eeq
Here on notes that the non-compact Hilbert space $M$ is modified into the non-compact $M^{H}$ due to the Hopf term in the $CP^{1}$ model.

In conclusion, we have studied the $CP^{1}$ model possessing the Hopf term which respects the second class constraints and 
is related with the BRST operator $Q$. Exploiting $Q$, we have explicitly constructed its de Rham cohomology group by deriving the ensuing quotient group via both the collections of all $Q$-closed and $Q$-exact ghost number $p$-forms. Moreover, we have introduced the $CP^{1}$ model without the Hopf term to investigate the ensuing effect of the Hopf term on the cohomology group. We have concluded that the Hopf term effects on the de Rham cohomology originate from the non-compact Hilbert space modified by this Hopf term. This is one of main results of the algebraic topology aspects of the $CP^{1}$ model with the Hopf term.

\end{document}